\documentclass[useAMS,usenatbib]{mn2e}

\newcommand{\uk}{{\rm $\mu$K\,} }
\newcommand{\dg}{$^\circ$}

\title[Methods for the Pointing Reconstruction of the {\it Planck} satellite]{Methods for the Pointing Reconstruction of the {\it Planck} satellite}
\author[D. L. Harrison]{D. L. Harrison$^{1,2}$\thanks{E-mail: dlh@ast.cam.ac.uk (DLH)}, F. van Leeuwen$^{2}$ and M. Rowan-Robinson$^{1}$\\
$^{1}$ Astrophysics Group, Blackett Laboratory, Imperial College, Prince Consort Road, London, SW7 2BW, UK\\
$^{2}$ Institute of Astronomy, Madingley Road, Cambridge, CB3 0HA, UK\\}
\begin{document}

\date{}

\pagerange{\pageref{firstpage}--\pageref{lastpage}} \pubyear{2003}

\maketitle

\label{firstpage}

\begin{abstract}
  This paper presents a method to reconstruct the position of the line of sight and orientation of the focal plane of the {\it Planck} satellite, which may be expressed by the parameters of the effective boresight and roll angles, respectively. The accuracy to which these parameters must be determined to avoid compromising the reconstruction of the $C_{\ell}$ power spectrum is assessed. It is expected that the pointing reconstruction will be performed using point source transits in the highest frequency channels, given their smaller beam sizes. Estimates are made for the number of galactic and extragalactic point sources visible by {\it Planck} in these channels. The ability of the method presented here to reconstruct the pointing parameters to the required accuracies using these point sources is investigated. It is found that the scanning strategy employed and the actual orientation of the focal plane both independently influence the available number of suitable point sources, from which this method may successfully evaluate the pointing parameters.  While this paper focuses on the pointing reconstruction of the {\it Planck} satellite, elements of the analysis presented here may be of use elsewhere.

\end{abstract}


\section{Introduction}

{\it Planck} is a European Space Agency satellite designed to produce high-resolution temperature and polarisation maps of the CMB. It possesses detectors sensitive to a wide range of frequencies from 30 to 857 GHz, split between two instruments the HFI and LFI, the high and low frequency instruments, respectively. {\it Planck} is scheduled for launch in 2007, when it will be inserted into a Lissajous orbit around the second Lagrange point of the Earth-Sun system. {\it Planck} will spin about its axis once per minute, sweeping its detectors along an almost great circle, the line of sight being almost perpendicular to the spin axis. The spin axis will nominally be repositioned every hour, and the roughly 60 or so circles corresponding to a single spin axis positioning may be binned together to form a ring. The spin axis passes through the centre of the solar panels and is directed towards the sun, thus maintaining the rest of the satellite in a cone of shadow produced by the solar panels.  The scanning strategy is determined by the frequency and the positions of the spin axis repointings over the course of the mission. The baseline scanning strategy is such that the spin axis remains in the ecliptic plane throughout the course of the mission and it maintains its sun-wards direction by hourly respositionings of 2.5\arcmin. In reality this scanning strategy will not be employed due to the lack of coverage at the ecliptic poles; the actual scanning strategy, however, has yet to be finalised.

This paper will investigate some of the issues concerning the evaluation of parameters required for the pointing reconstruction of the {\it Planck} satellite, specifically those resulting from non-zero off-diagonal elements in the inertia tensor as defined in the nominal satellite reference frame. The slight misalignment between the nominal spin axis of the satellite and the nearest principal axis of the inertia tensor defines the effective boresight angle and the rotation of the focal plane. These parameters may only be established using the science data, and they will vary over the course of the mission due to changes in the inertia tensor as consumables are depleted. It is therefore important to assess what the accuracy requirements on these parameters are, and whether it will be possible to extract the parameters to those required accuracies on a sufficiently regular basis.

 The variation of the misalignment over the course of the mission is not yet well defined and is currently under investigation. It is, however, expected that the frequency at which the roll and effective boresight angles must be evaluated will be of the order of once a week \cite{puget01}. 

In Section~\ref{section_acc_req} the accuracy requirements of the effective boresight angle and the roll angle, the rotation of the focal plane, are discussed. Methods of evaluating the roll and effective boresight angles through the use of galactic and extragalactic point sources are discussed in Section~\ref{section_methods}. The number of both galactic and extragalactic point sources likely to be detectable by {\it Planck} are discussed in Section~\ref{pt_sources_section}. The simulated data on which our methods are tested is discussed in Section~\ref{section_simData}, the results of which are discussed in Section~\ref{section_results}.

\section[]{Accuracy requirements}
\label{section_acc_req}

\begin{figure}
\begin{center}
\setlength{\unitlength}{1cm}
\begin{picture}(7,7)(0,0)
\put(8.5,0){\includegraphics{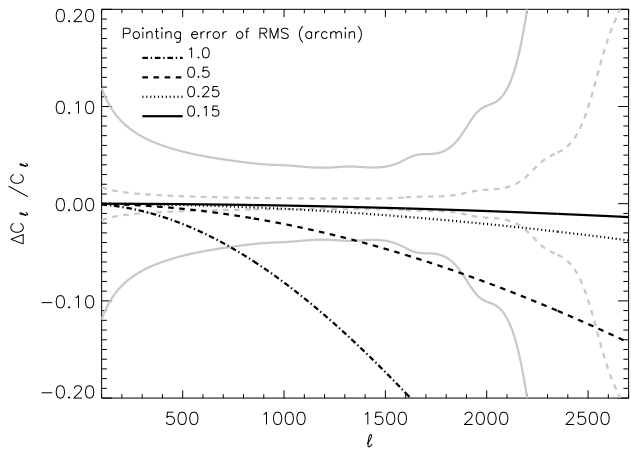}}
\end{picture}
\end{center}
\caption[]{The figure shows the unsubtractable errors on the $C_{\ell}$s which are dominated by cosmic variance to the left and the finite beam size to the right. The solid grey curves enclose the region corresponding to the error on each individual multipole, whereas the dashed grey curves correspond to the error on multipole bins of width 50. The black curves correspond to the additional errors on the reconstruction of the $C_{\ell}$s due to unknown random pointing errors. A beam of FWHM 5\arcmin, with a noise of 10.3\uk per beam and a fractional sky coverage of 0.7 have been assumed.}
\label{acc_fig}
\end{figure}

The effect of unknown random pointing reconstruction errors on the recovered ${\rm C_{\ell}}$ values may be seen in Figure~\ref{acc_fig}. This figure has been replicated from one first shown by~\cite{lamarre01}, using equation~\ref{c_ell_eqn} modified from~\cite{knox95}: 
\begin{equation}
\label{c_ell_eqn}
\frac{\Delta C_{\ell}}{C_{\ell}}=\sqrt{ \frac{2}{(2\ell+1) \times f_{sky}}} \left( 1 + \frac{\omega}{C_{\ell} W_{\ell}} \right)
\end{equation} where $\omega$ is the noise per beam, $f_{sky}$ is the fractional sky coverage and $W_{\ell}$ is the window function. The sensitivity of an experiment to the $\ell$ multipoles is given by its window function. For a map made with a Gaussian beam, the window function is given by: $W_{\ell}={\rm e}^{-\ell^2 \sigma_b^2}$. If the beam sigma, $\sigma_b$, is known, the decreased sensitivity to the higher $\ell$ multipoles may be accounted for by multiplying by the window function. It is this which results in the exponential increase in the noise for the high multipoles in Figure~\ref{acc_fig}, as the noise is undiminished by the beam size. For low multipoles the noise is dominated by the first term in equation~\ref{c_ell_eqn} which expresses the sample and cosmic variance.

Figure~\ref{acc_fig} uses the expected beam size and noise levels for the 217 GHz frequency channel and assumes 70 per cent sky coverage. The grey curves enclose regions corresponding to the unsubtractable noise, given by equation~\ref{c_ell_eqn}. The solid grey curves correspond to the error on each individual multipole, while the dashed grey curves the error on multipole bins of width 50.

The black curves, in Figure~\ref{acc_fig}, correspond to the additional errors in the reconstruction of the  $C_{\ell}$s due to the effects of unknown random pointing errors. This results in a decreased sensitivity to the higher $\ell$ multipoles as is seen in Figure~\ref{acc_fig}. The pointing uncertainty results in an effective smearing of the beam and hence an increase in the effective beam width. If this is well known, a new window function may be evaluated and the effects of the pointing errors will not be as dramatic as those shown.  However, in reality the pointing errors will be correlated, smearing the beam more in one direction, which will lead to asymmetric effective beams. The effect of smearing the beam predominately in one direction is to increase the sensitivity to higher $\ell$ multipoles as compared to the uniform smearing shown inFigure~\ref{acc_fig}. In other words, for a given magnitude, correlated pointing errors will lead to a smaller reduction in sensitivity at high $\ell$, than random pointing errors. However, given that the extent of the correlation of the pointing errors is not known and that the reduction in the maximum allowable pointing error is only significant for highly correlated pointing errors, the pointing errors are assummed to be random in the subsequent analysis.

Ideally, the pointing accuracy should be such that the additional errors introduced are less than the unsubtractable noise.

For each frequency channel, with the exception of 857 GHz in which the $C_{\ell}$s are not observable, the unsubtractable errors on the $C_{\ell}$s, given the expected beam size and noise levels, were evaluated. Hence, the ideal pointing accuracy required for each frequency channel may be found. These are shown in Table~\ref{pt_acc_table}, for the cases of 70 and 100 percent sky coverage.

\begin{table}

\caption{The requirement on the random pointing error versus frequency channel, given fractional sky coverage,  $f_{sky}$.}
\label{pt_acc_table}

\begin{tabular}{|ccccc|}
\hline
Frequency & FWHM & noise per & \multicolumn{2}{c}{maximum random}  \\
Channel & &  beam & \multicolumn{2}{c}{pointing error} \\
 & & & $f_{sky}=0.7$ & $f_{sky}=1.0$ \\
 (GHz) & (\arcmin) & (\uk) & (\arcmin) & (\arcmin) \\ 
\hline
100 & 9.2 & 6.0 & 0.23 & 0.21\\
143 & 7.1 & 6.5 & 0.18 & 0.16 \\
217 & 5.0 & 10.3 & 0.16 & 0.15 \\
353 & 5.0 & 40.9 & 0.29 & 0.26 \\
545 & 5.0 & 46.3 & 0.31 & 0.28 \\
\hline
\end{tabular}
\end{table}

The uncertainty in the roll angle should be such that the corresponding uncertainty in the position of the detector is less than the requirement set by the analysis shown in Table~\ref{pt_acc_table}. The required accuracy of the roll angle is determined by a combination of the maximum allowable pointing error and the location of the detectors in the focal plane. The further from the centre of the focal plane the more the detectors will be affected by the roll angle. The required accuracy of the roll angle for each frequency channel is displayed in Table~\ref{roll_acc_table}, again for both the cases of 70 and 100 percent sky coverage.

\begin{table}

\caption{The requirement on the error in the roll angle, given fractional sky coverage, $f_{sky}$.}
\label{roll_acc_table}

\begin{tabular}{|ccc|}
\hline
Frequency Channel& \multicolumn{2}{c}{Roll Angle Error} \\
 & $f_{sky}=0.7$& $f_{sky}=1.0$ \\
 (GHz) & (\arcmin) & (\arcmin)\\ 
\hline
100 & 6.9 & 6.3\\
143 & 4.9 & 4.4\\
217 & 6.7 & 6.2\\
353 & 8.1 & 7.2\\
545 & 8.3 & 7.5\\
\hline
\end{tabular}
\end{table}
 
As shown in Table~\ref{roll_acc_table}, the 143 GHz channel sets the requirement on the uncertainty in the roll angle to be 4.4\arcmin; assuming no other pointing errors from other sources.

\section[]{Methods}
\label{section_methods}

\begin{figure}
\begin{center}
\setlength{\unitlength}{1cm}
\begin{picture}(8,8)(0,0)
\put(9,0){\includegraphics{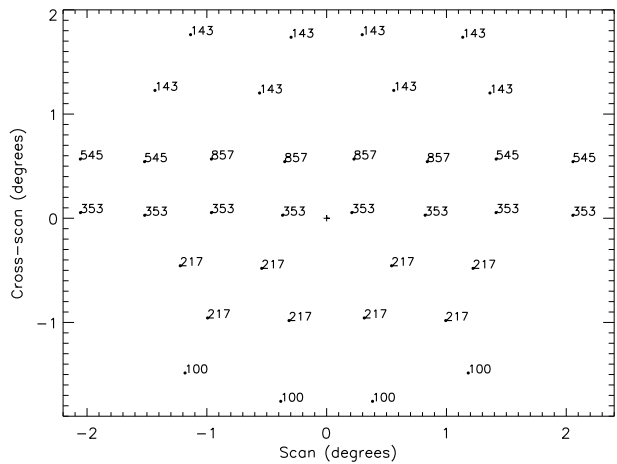}}
\end{picture}
\end{center}
\caption[]{This figure displays the focal plane layout of the HFI, in terms of the nominal scan and cross-scan directions. The positive scan direction indicates the direction in which the focal plane traverses the sky. A source will therefore cross the focal plane from right to left in this figure. The exact location of the detectors are given by the small filled circles. The top four frequency channels have offset detectors to ensure full sampling in the cross-scan direction. The centres of adjacent detectors are offset by 1.5\arcmin\, in the cross-scan direction, with every detector being in a pair with the same cross-scan location. The roll angle is defined as a rotation about the FRP, which is shown by the cross, and the effective boresight is defined as the angle between the spin axis and the centre of the focal plane. In this paper we will define a positive roll angle as a clockwise rotation, and a negative rotation as any anti-clockwise rotation about the FRP.}
\label{focal_fig}
\end{figure}

In this section methods for evaluating the roll angle and the effective boresight, using measurements of point sources, will be considered. The effect of a roll angle will be to change the positions of the detectors with respect to the scan and cross-scan directions, so any method to evaluate the roll angle will depend on the focal plane layout. Figure~\ref{focal_fig} illustrates the layout of the focal plane. The four highest frequency channels have offset detectors to ensure full sampling in the cross-scan direction. The centres of adjacent detectors are offset by 1.5\arcmin\, in the nominal cross-scan, with every detector being in a pair with the same nominal cross-scan location. The highest frequency channels will be the most suited to extracting the roll angle and effective boresight angle, due to the smaller beam size and the larger numbers of point sources which will be detected at the higher frequencies. For this paper, unless otherwise stated, we will consider only the 857 GHz detectors. We will also define a positive roll angle as a clockwise rotation, and a negative rotation as any anti-clockwise rotation about a fiducial reference point, FRP, in the centre of the focal plane.

\begin{figure}
\begin{center}
\setlength{\unitlength}{1cm}
\begin{picture}(11,8)(0,0)
\put(-1.8,11){\includegraphics{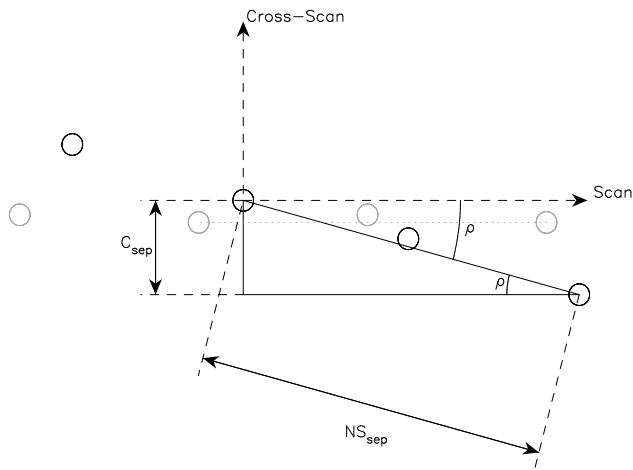}}
\end{picture}
\end{center}
\caption[]{This figure shows the effect of a change in the orientation of the focal plane with respect to the scan direction. The dotted grey circles correspond to detector positions in the nominal orientation of the focal plane, whereas the solid black circles correspond to their positions after a rotation, $\rho$, about the FRP. For the highest frequency channels the centres of alternate detectors are aligned in the nominal scan direction as shown in the figure by the dotted grey line. Any difference in the cross-scan position of these detectors is therefore indicative of a non-zero focal plane rotation. The nominal scan separation, ${\rm NS_{sep}}$, of alternate detectors may be used with their cross-scan separation, ${\rm C_{sep}}$, to calculate the roll angle.}
\label{def_axes_fig}
\end{figure}

In the nominal case of zero roll angle there are two pairs of detectors with the same cross-scan position. Any separation found between the cross-scan positions of these detectors may be used to infer the roll angle, as shown in Figure~\ref{def_axes_fig}.

 In order to evaluate the roll angle a method is needed to find the cross-scan positions of the detectors, or the cross-scan separation of the detectors aligned in the nominal scan direction, directly from the amplitudes of a point source as observed by each detector. Once a method has been found it may be investigated as to what signal-to-noise ratio, SNR, is required to obtain the roll angle to the required accuracy and hence an estimate of the number of available point sources may be made. Methods for reconstructing the roll angle are discussed in Section~\ref{subsec_roll_angle}.

The effective boresight is the angle between the spin axis and the scan circle which passes through the FRP, shown as $\alpha$ in Figure~\ref{sphere_fig}. In order to evaluate the effective boresight the cross scan position of a known point source relative to the FRP must be found. Once the roll angle has been evaluated, and hence the cross scan positions of the detectors are known, the evaluation of the cross scan positions of point sources becomes possible. Methods for evaluating the cross scan positions of point sources will be discussed in Section~\ref{section_eff_boresight}.

These methods are applied to data from one or two pointings of the spin axis, as required. The data from each spin axis pointing may be used in terms of the individual revolutions, circles, or the accumulated revolutions corresponding to the spin axis pointing, a ring. In the text below a scan circle may be thought of as the path of the line of sight of the FRP on the sky in the case of a circle or the average path in the case of a ring. This differentiation is necessary as in reality the spin axis describes a small rotating ellipse on the sky. This nutation of the spin axis results in deviations from the expected path of the line of sight given a spin axis rotating around a fixed point. These deviations are small and will average out in the ring data as the period of the nutation is greater than the rotation rate of the satellite. Providing that the amplitude of the nutation remains small it may be safely ignored in all cases, should this no longer be the case the effects of the nutation may be corrected for using information on the nutation parameters as found from the Star Tracker data. Throughout this paper it is assumed that these methods will be applied to rings, given their lower noise levels and their greater robustness to the effects of nutation.

\subsection{Reconstructing the Roll Angle}
\label{subsec_roll_angle}

\subsubsection{Using a single ring}
\label{subsec_one_ring}

\begin{figure}
\begin{center}
\setlength{\unitlength}{1cm}
\begin{picture}(8,8)(0,0)
\put(9,0){\includegraphics{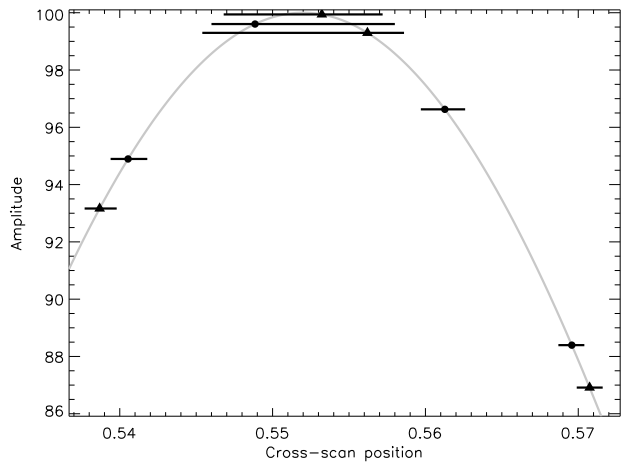}}
\end{picture}
\end{center}
\caption[]{This figure displays the amplitudes, as observed by each detector, for two different roll angles, given a point source with an amplitude of 100 times the 1$\sigma$ noise and  cross-scan position from the FRP of 0\fdg552. The circles represent the detected amplitudes given a roll angle of 0\fdg4, while the triangles 0\fdg7. The 1$\sigma$ errors in the amplitudes have been converted into errors in the cross-scan position of the detectors. The corresponding 1$\sigma$ uncertainties in the positions of the detectors overlap, it is therefore not possible to establish the true roll angle without a higher signal-to-noise point source.}
\label{degen_fig}
\end{figure}

In the first instance one may consider evaluating the roll angle using a single ring, using the observed amplitudes from the detectors to solve for the amplitude of the point source and its cross-scan position. Here a ring consists of the data corresponding to a single repointing of the spin axis, and the observed amplitude and location in the scan direction of the point source is found for each detector from either binning or by phase-ordering~\citep{leeuwen01} this data. Once the amplitude and cross-scan position of the point source has been evaluated the amplitude observed by each detector may be used to evaluate its cross-scan separation from the point source position. Given these relative cross-scan locations the roll angle may be evaluated. However, in practice the errors in the detected amplitudes result in errors in the cross-scan locations which may lead to a large uncertainty in the recovered roll angle. An example of this is shown in Figure~\ref{degen_fig}, in which the errors in the detected amplitude have been converted into errors in the cross-scan position of the detector. In this figure it has been assumed that the amplitude and cross-scan position of the point source have been successfully recovered with negligible error; in this example the amplitude of the point source is 100 times the 1$\sigma$ noise and its cross-scan position from the scan circle which contains the FRP is 0\fdg552. For comparison two scenarios are plotted; the circles represent the detected amplitudes in the four detectors resulting from a 0\fdg4 roll angle, while the triangles correspond to the amplitudes resulting from a 0\fdg7 roll angle. The 1$\sigma$ errors in the amplitudes have been converted into errors in the cross-scan position of the detectors. Working across the figure pairing up the adjacent triangles and circles which correspond to the same detector, it is possible to see that uncertainties in the cross-scan positions, for each of the roll angles, overlap for all of the detectors. Hence in this example it is not possible to successfully distinguish between roll angles of 0\fdg4 to 0\fdg7. Other cross-scan positions of the point source with respect to the FRP result in similar difficulties over different ranges of roll angle.

In reality there will be errors in the recovered amplitude and cross-scan position of the point source, together with errors in the relative calibration between detectors which will increase the effective errors in the cross-scan positions of the detectors and hence the uncertainty in the recovered value of the roll angle. Indeed, it is not possible to evaluate the cross-scan position of the point source, without first knowing the value of the roll angle due to the lack of knowledge on the positions of the detectors.

\subsubsection[]{Using Two Rings}
\label{two_rings}

\begin{figure}
\begin{center}
\setlength{\unitlength}{1cm}
\begin{picture}(10,10)(0,0)
\put(0,0){\includegraphics{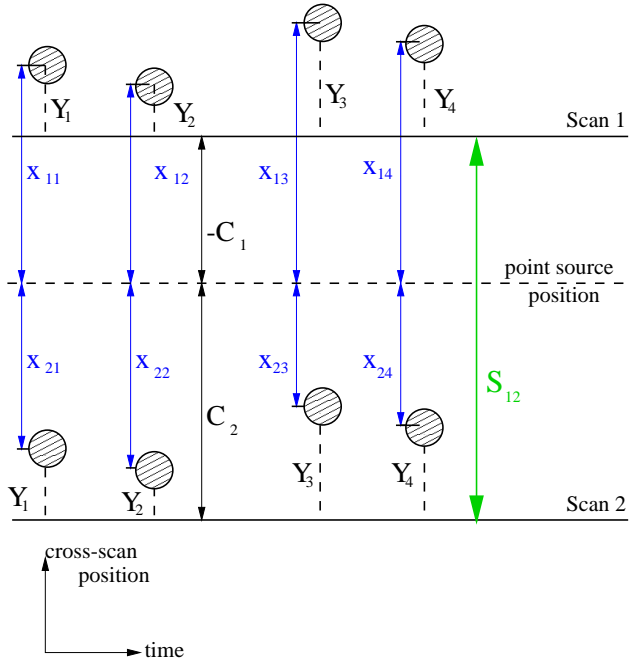}}
\end{picture}
\end{center}
\caption[]{This figure shows two neighbouring, although not necessarily adjacent scans, which pass on either side of a bright point source. The detector positions relative to the scan are denoted by $Y_{j}$, while the $X_{ij}$ represents the offset from the point source of detector j in scan i. $C_{i}$ is the cross-scan position of the point source from scan i and $S_{12}$ is the distance between the two scans at the location (longitude) of the point source.}
\label{roll_fig}
\end{figure}

\begin{figure}
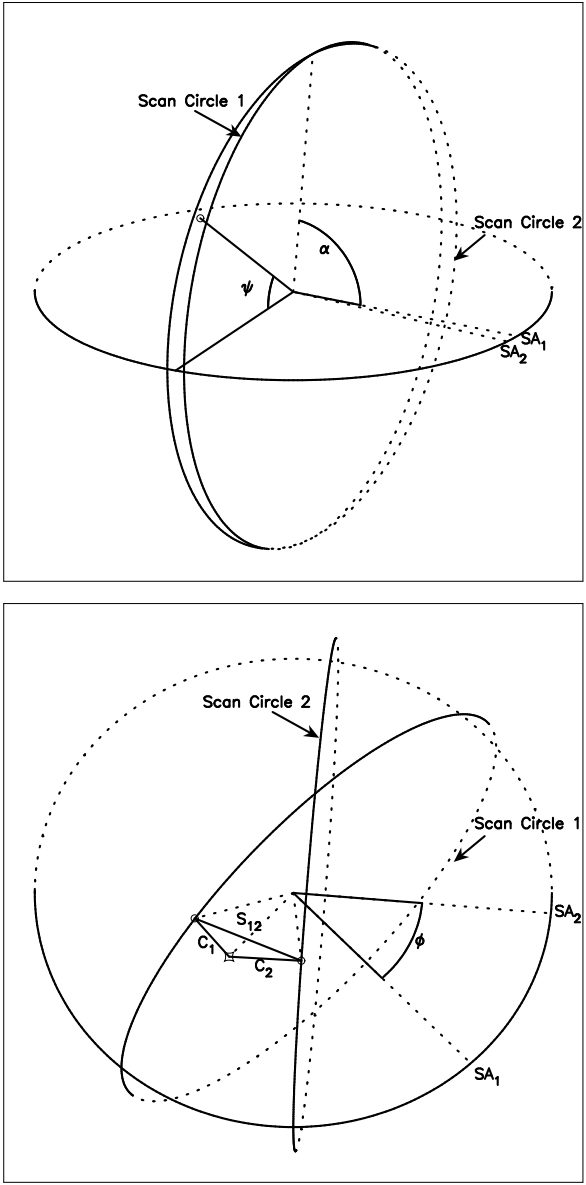

\begin{center}
\setlength{\unitlength}{1cm}
\begin{picture}(10,16)(0,0)
\put(-1.5,18){\includegraphics{sphere_fig1.ps}}
\put(-1.5,10){\includegraphics{sphere_fig2.ps}}
\end{picture}
\end{center}
\caption[]{This figure shows the angles discussed in the text relating to two neighbouring scan circles with a point source located between them. The top diagram shows the plane in which the motion of the mean spin axis has occurred, and the angle between this plane and the point source, $\psi$. The angle between the spin axis and the scan circle, $\alpha$, is the effective boresight. The angle $\psi$, may be evaluated once the reference phase, or time of first transit of the FRP is known, as described in \cite{leeuwen01}. In the first instance this may be evaluated by the Star Trackers. The lower diagram shows the angle, $\phi$, through which the spin axis has been repositioned. The angles $C_1$ and $C_2$ represent the cross-scan angular distance between the point source and scan circles 1 and 2, respectively. The angle $S_{12}$ is the angular separation between the intersection points of $C_1$ and $C_2$ on scan circles 1 and 2. The angle $\phi$, between the two spin axis positions, has been exaggerated for clarity in this figure; in the analyis presented here $\phi$ never exceeds 1\dg. }
\label{sphere_fig}
\end{figure}

If there are no abrupt changes in the inertia tensor during the mission, such as fuel shifts, the roll and effective boresight angles will only change slowly over the duration of the mission. Assuming that this is indeed the case and hence the roll angle is a slowly varying parameter, its evaluation may be improved by including the additional information available from the detection of the same point source on a neighbouring ring. Figure~\ref{roll_fig} shows schematically the relative locations of the detectors to the scan directions, corresponding to the two rings, and the point source position. The cross-scan position of the point source from scan 1 and scan 2 is given by  $C_1$ and $C_2$, respectively; $C_1$ has a negative sign so that the sign convention for the cross-scan position may be the same for both scans. As the roll angle is unchanged, the cross-scan positions of the detectors, $Y_{j}$, from the FRP, will therefore not have changed between the two neighbouring scans. The sum of the offset from the point source position of a detector in scan1, $x_{1j}$, and the same detector in scan2, $x_{2j}$ will therefore be equal to the sum of the angles $C_1$ and $C_2$. The angular separation between the two scans at the location of the point source, $S_{12}$, may also be related to the sum of the angles $C_1$ and $C_2$, however in this case this relationship is only an approximate. Though as will be shown below for neighbouring scans, it may be thought of as being exact:
\begin{equation}
\label{setup1}
S_{12}\approx-C_1+C_2.
\end{equation}	
The angle between the two scans, $S_{12}$, at the location of the point source depends on the separation of the mean spin axis positions corresponding to each of the scans, and the elevation, $\psi$, of the point source from the plane that connects the mean spin axis positions.  Figure~\ref{sphere_fig} shows a schematic of the two scans and the point source in question. The separation between the mean spin axis positions is given by $\phi$, this is also the maximum separation between the two scans. The angles, $\psi$ and $\phi$, may be evaluated using additional information provided by the Star Trackers. In the case of $\phi$, it may be evaluated from the two recovered mean spin axis positions. The evaluation of $\psi$ also requires the time of first transit of the FRP though the reference circle which connects the mean spin axis position and the North Ecliptic Pole, NEP, as defined in \cite{leeuwen01}.

The angle $S_{12}$, also has a dependence on the effective boresight, $\alpha$, though given the expected value, $\alpha \approx $ 85\dg, this dependence is weak:
\begin{eqnarray}
\label{s12_eqn}
\cos(S_{12}) & = & \cos(\phi) \left(\cos^{2}(\alpha)+\sin^{2}(\alpha)\cos^{2}(\psi)\right) \nonumber \\
& & \mbox{} + \sin^{2}(\alpha)\sin^{2}(\psi).
\end{eqnarray} As can be seen from Figure~\ref{sphere_fig} using the value of $S_{12}$ for the sum of the two cross-scan angles to the point source, $-C_1+C_2$, will result in an underestimation of this value. This underestimation will be the greatest when the point source lies midway between the two scans. To quantify the magnitude of this underestimation, the angle between the two cross-scan directions, $\beta$, given a point source which lies midway between the two scans may be found using:
\begin{eqnarray}
\label{beta_eqn}
\cos(\beta) & = & \cos(\pi-\phi) \left(\cos^{2}(\psi)\cos^{2}(\alpha)+\sin^{2}(\alpha)\right) \nonumber \\
& & \mbox{} - \cos^{2}(\alpha)\sin^{2}(\psi).
\end{eqnarray} Given that we are only interested in neighbouring scans the angle, $\phi$ will be small and the angle, $\delta$, between $ C_{1,2}$ and $ S_{12}$ may then be found using:
\begin{equation}
\label{delta_eqn}
\delta=\frac{\pi-\beta}{2}.
\end{equation}	For $\phi \le$ 1\dg, which corresponds to  $\le$ 24 nominal repointings of 2.5\arcmin,  the underestimation in the sum of the cross-scan angles is of the order of a thousandth of a percent. This corresponds to an overestimation for the magnitudes of the roll angle considered here of the order of a hundredth of an arc-second. Hence, for our present purposes the relation in equation~\ref{setup1} may be considered to be exact.
  
From Figure~\ref{roll_fig} it is seen that the sum of the cross scan angles $C_1$ and $C_2$ is equal to the sum of the offsets, from the point source position, of the ${\rm j^{th}}$ detector in scans 1 and 2, $x_{1j}$ and $x_{2j}$. The angle $S_{12}$ is therefore also equal to the sum of these offsets, via equation~\ref{setup1}:
\begin{equation}
\label{setup2}	
S_{12}=x_{1j}+x_{2j}.
\end{equation}	The cross-scan position of the point source from the scan and the offsets of the detectors from the point source, may be related to the position of the detectors relative to the scan:
\begin{equation}	
\label{setup3}	
-C_1=x_{1j}-Y_j	,
\end{equation}	
\begin{equation}	
\label{setup4}	
C_2=x_{2j}+Y_j	.
\end{equation}	The observations of the point source may be described by the following, the amplitude of the point source as seen by the ${\rm j^{th}}$ detector in the ${\rm i^{th}}$ scan is denoted by  $A_{ij}$:
\begin{equation}
A_{ij}=A \exp \left( \frac{-x_{ij}^2}{2 \sigma_{j}^2} \right).
\end{equation} Where $\sigma_{j}$ is the $1\sigma$ width of the assumed Gaussian beam for the ${\rm j^{th}}$ detector.\\

The expression for the ratio of the amplitudes of the point source as observed in the two scans by the ${\rm j^{th}}$ detector may be rearranged to find:
\begin{equation}
x_{2j}^2-x_{1j}^2=2\sigma_{j}^2 \ln \left( \frac{A_{1j}}{A_{2j}} \right),
\end{equation} from which, using equation~\ref{setup2}, we find:
\begin{equation}
\label{shows_adv_eqn}
x_{2j}-x_{1j}=\frac{2\sigma_{j}^2}{S_{12}} \ln \left( \frac{A_{1j}}{A_{2j}} \right), 
\end{equation} where everything on the RHS of this equation is known observationally. Equation~\ref{shows_adv_eqn} shows a major benefit of this method; as it is based on the ratio of the amplitudes seen by the same detector in different scans, uncertainties in the relative calibrations between detectors will be irrelevant.\\

An alternative expression for the difference between the offsets of the ${\rm j^{th}}$ detector from the point source on each scan, may be found using equations~\ref{setup3} and~\ref{setup4}:
\begin{equation}
\label{observables}	 
Y_j= \frac{1}{2} \left((C_2+C_1)- (x_{2j}-x_{1j}) \right).
\end{equation}	

The cross-scan separation, ${\rm C_{sep}}$, of the aligned detectors may be found:
\begin{equation}
Y_3 -Y_1 = \frac{1}{2} \left( (x_{21}-x_{11})-(x_{23}-x_{13}) \right) ,
\end{equation}

\begin{equation}
Y_4 -Y_2 = \frac{1}{2} \left( (x_{22}-x_{12})-(x_{24}-x_{14}) \right) ,
\end{equation}

\begin{equation}
{\rm C_{sep}}=Y_3 -Y_1 = Y_4 -Y_2 ,
\end{equation} and hence the roll angle, {$\rm \rho$}, may be evaluated, where the denominator, ${\rm NS_{sep}}$, is determined by the separation in the scan direction of the aligned detectors in the case of zero roll angle, and is shown in Figure~\ref{def_axes_fig}:

\begin{equation}
{\rm \sin(\rho)} =\frac{\rm C_{sep}}{\rm NS_{sep}}.
\end{equation}

\subsection{Evaluating the Effective Boresight}
\label{section_eff_boresight}

The effective boresight, which is the angle between the mean spin axis position and the scan circle, may be found if a point source of known position is observed in the data from the scan circle. The cross scan position, C, of the point source is found with respect to the FRP, and hence the scan circle. Here the cross scan position of the point source is defined as positive if the point source lies on the opposite side of the scan circle to the spin axis position. The effective boresight, $\alpha$, may then be found using:
\begin{equation}
\label{eff_B_eqn2}
\alpha = \omega - C,
\end{equation}	where $\omega$ is the angle between the point source position,  $(\beta_{pt}, \lambda_{pt})$, and the mean spin axis position, $(\beta_{sa}, \lambda_{sa})$, and is given by:
\begin{equation}
\label{eff_B_eqn}
\cos \omega = \cos \beta_{pt}  \cos \beta_{sa} \cos(\lambda_{pt} - \lambda_{sa}) + \sin \beta_{pt} \sin \beta_{sa}.
\end{equation}	

The cross scan position of a point source may be found using data from a single ring or by an extension of the two ring method for the evaluation of the roll angle. 

In the case of data from a single ring, the amplitudes found from the transits of the point source are fit by a Gaussian beam profile to establish best fit values for the position and amplitude of the peak. This method will only provide unbiased cross scan position for the point source if it lies between the range of cross scan positions of the detectors. Hence the value of the roll angle is required prior to the evaluation of the cross-scan position of the point source. Additionally, uncertainties in the relative calibration and/or beams of the detectors will increase the uncertainty in the recovered value of the cross-scan position and hence effective boresight angle.

Alternatively, by an extension to the method to extract the roll angle from two rings, the cross-scan position of the point source may be found using the ratio of the amplitudes observed by each detector in two neighbouring scans.

The sum over all detectors using equation~\ref{observables}, may be used to evaluate the difference between the cross-scan positions of the point source to each scan:
\begin{equation}
\label{cross_dif_eqn}	
(C_2+C_1) = \frac{1}{N} \sum^N_j \left( x_{2j}-x_{1j} \right) + \frac{2}{N} \sum^N_j Y_j ,
\end{equation}	where N=4 using a single frequency channel, or N=8 combining both the 857 and 545 GHz channels.

Once the roll angle is found the positions of the detectors, $Y_j$, are known, indeed the sum over the positions of the detectors is approximately a constant for small values of the roll angle. The cross scan position of the point source may be found for each scan by combining equations~\ref{setup1} and~\ref{cross_dif_eqn}:
\begin{equation}
\label{cross1_eqn}
C_1 = - \frac{S_{12}}{2}+ \frac{1}{N} \sum^N_j \left( \sigma_{j}^2 \ln \left( \frac{A_{1j}}{A_{2j}} \right) \right) +  \frac{1}{N} \sum^N_j Y_j  ,
\end{equation}
\begin{equation}
\label{cross2_eqn}
C_2 = \frac{S_{12}}{2}+ \frac{1}{N} \sum^N_j \left( \sigma_{j}^2 \ln \left( \frac{A_{1j}}{A_{2j}} \right) \right) +  \frac{1}{N} \sum^N_j Y_j .
\end{equation}

Once the cross scan position of the point source is known with respect to the FRP, the effective boresight may be evaluated via equations~\ref{eff_B_eqn2} and~\ref{eff_B_eqn}, if the position of the point source is known.

\section{Predicting the number of point sources visible with {\it Planck} HFI}
\label{pt_sources_section}

The numbers of point sources visible with {\it Planck} HFI may be predicted by using the IRAS point source catalogue (PSC, \cite{beichman88}). This is a catalogue of some 250,000 well-confirmed point sources, providing positions, flux densities at 12, 25, 60 and 100\micron , uncertainties and various cautionary flags which are given for each source. Only the brightest of these sources will be visible by {\it Planck}. Since the brightest of these sources have an increased chance of actually being extended objects, it is important to establish the characteristics which best indicate a point source nature. In order to establish these characteristics a population of objects which are known to be point sources is required. Such a population of point sources may be found by applying the following stellar colour cuts \citep{mrr86} to the PSC:

\begin{eqnarray}
\log \left( \frac{S_{100}}{S_{60}} \right) & \leq  & -0.3 , \nonumber\\
\log \left( \frac{S_{60}}{S_{25}} \right)  & \leq  & -0.3 ,\nonumber\\
\log \left( \frac{S_{25}}{S_{12}} \right)  & \leq  & 0.2 .
\end{eqnarray}

The resultant population of stellar, and hence point sources may then be investigated. The population consists of 450 bright, ${S_{100}} \geq 2.0$ Jy, point sources which were all found to have a 60\micron\, correlation coefficient of grade A, where the correlation coefficient is an indication of how closely the data match the template for response from a point source. The 100\micron\, correlation coefficient was found to be grade A for 331 of these sources, grade B for 104, and grade C for 12. The reliable indicators of a point source were then taken to be a 60\micron\, correlation coefficient of grade A and a 100\micron\, correlation coefficient of grade A or B. 

The above cuts on the correlation coefficients were applied to the IRAS PSC, excluding those sources with correlation coefficients worse than grade A at 60\micron\, and grade B at 100\micron. In addition sources with only upper limits at 100 or 60\micron\, were also excluded. The colour cut $\log \left( \frac{S_{100}}{S_{60}} \right) >  -0.3 $ was applied, in order to remove all the sources with rapidly declining fluxes towards longer wavelengths. After these cuts the number of sources remaining is 7874, 5588 of which are in the PSCz catalogue \citep{saunders00}, the rest predominately populate the galactic plane.

The IRAS galaxies may be modelled using the work of \cite{dunne00} and \cite{dunne01} who made 850\micron\, and 450\micron\, maps of galaxies from the IRAS Bright Galaxy Sample. \cite{dunne01} find that this combined data is best fit by a two component spectral energy distribution, SED. The two components are modelled as two $\beta =2$ grey bodies corresponding to warm and cold dust temperatures:
\begin{equation}
S_{\nu}= N_w \nu^{\beta} B(\nu,T_w) + N_c \nu^{\beta} B(\nu, T_c),
\end{equation} where B($\nu$,T) is the Planck function for both components, $T_w$ and $T_c$ are the temperatures of the warm and cold components and $N_w$ and $N_c$ correspond to the relative masses of the dust in each component. 

The extragalactic point source fluxes may now be extrapolated to {\it Planck} frequencies. Since a two component dust model is a four parameter fit, two temperatures and two normalisations, and as the 12\micron\, flux is only present as an upper limit for some of these sources, the average cold temperature found by \cite{dunne01} of  20.9K is used and the remaining three parameters are solved for using the 25, 60 and 100\micron\, fluxes. Using these parameters the fluxes of each galaxy at 350, 550 and 850\micron\,, corresponding to the {\it Planck} frequencies of 857, 545 and 353 GHz, are found. 

\begin{table}

\caption{Number of extragalactic point sources above a flux which corresponds to a given SNR in a binned ring}
\label{exgal_pred_table}

\begin{tabular}{rrrrrrr}
\hline
SNR & \multicolumn{6}{c}{Frequency (GHz)}\\
 & \multicolumn{2}{c}{857} & \multicolumn{2}{c}{545} & \multicolumn{2}{c}{353} \\
 & S (Jy) & N$\geqslant$S & S (Jy) & N$\geqslant$S & S (Jy) & N$\geqslant$S \\
\hline
10 & 2.84 & 452 & 2.51 & 82 & 1.54 & 26 \\
50 & 14.20 & 59 & 12.55 & 14 & 7.70 & 4 \\
100 & 28.40 & 22 & 25.10 & 7 & 15.40 & 1 \\
200 & 56.80 & 9 & 50.20 & 4 &  30.80 & 0 \\
400 & 113.60 & 5 & 100.40 & 0 & 61.60 & 0 \\
\hline
\end{tabular}
\end{table}

The number of point sources above a required flux level at each frequency may now be found. The flux levels chosen are those which correspond to a range of signal-to-noise ratios in a binned ring. This allows a direct comparison between frequency channels as each frequency channel is expected to have different noise levels.  Table~\ref{exgal_pred_table} shows the number of extragalactic point sources above a given signal-to-noise ratio with the corresponding flux for the top three {\it Planck} frequency channels.

The galactic point sources are modelled using a single grey body fit using $\beta=2.3$, as expected for compact HII regions, M-type stars and Carbon stars, to extrapolate to {\it Planck} frequencies:

\begin{equation}
S_{\nu}= \nu^{\beta} B_{\nu} =  \nu^{\beta} \frac{2h\nu^3}{c^2} \frac{1}{\exp(h\nu/kT)-1}.
\end{equation}

\begin{table}
\caption{Number of galactic point sources above a flux which corresponds to a given SNR in a binned ring}
\label{pred_gal_table}
\begin{tabular}{rrrrrrr}
\hline
SNR & \multicolumn{6}{c}{Frequency (GHz)}\\
 & \multicolumn{2}{c}{857} & \multicolumn{2}{c}{545} & \multicolumn{2}{c}{353} \\
 & S (Jy) & N$\geqslant$S & S (Jy) & N$\geqslant$S & S (Jy) & N$\geqslant$S \\
\hline
10 & 2.84 & 1147 & 2.51 & 575 & 1.54 & 269 \\
50 & 14.20 & 545 & 12.55 & 203 & 7.70 & 69 \\
100 & 28.40 & 361 & 25.10 & 122 & 15.40 & 33  \\
200 & 56.80 & 219 & 50.20 & 57 & 30.80 & 14 \\
400 & 113.60 & 131 & 100.40 & 25 & 61.60 & 4 \\
\hline
\end{tabular}
\end{table} Table~\ref{pred_gal_table} shows the number of galactic point sources above a given signal-to-noise ratio with the corresponding flux for the top three {\it Planck} frequency channels.

As seen in Tables~\ref{exgal_pred_table} and~\ref{pred_gal_table}, the vast majority of the point sources will be detected in the highest frequency channel of 857 GHz. Once the number of point sources above given signal-to-noise ratios have been found, it is possible to estimate the number of them from which the roll angle may be evaluated. Not all potential locations of a point source relative to two neighbouring scans, shown by Figure~\ref{roll_fig}, may successfully be used to extract the roll angle as will be examined below in Section~\ref{section_results}.

\section{Simulated Data}
\label{section_simData}

In order to test the performance of this method, simulated observations were generated. This data was comprised of the peak amplitudes of the transit of a point source for each of the four 857 GHz detectors in two neighbouring scans.

In generating the simulated observations, the location of the point source was fixed to lie on or between two successive scans, separated by a single repointing, shown schematically by the shaded area in Figure~\ref{area_sphere_fig}.  The signal-to-noise ratio of the point source, its position between the two scans and the angle between the point source and the plane of motion of the spin axes, $\psi$ in Figure~\ref{sphere_fig}, were varied for different roll angles. Henceforth the angle, $\psi$ will be referred to as phase. It should be noted that this definition of phase is different from that which appears in \cite{leeuwen01}. In the simulated observations it is assumed that the separation between successive repointings of the spin axis is the nominal 2.5\arcmin. The peak amplitudes of the transit of the point source are then found in two neighbouring scans, separated by a given number of repointings. In summary the simulated observations vary the following parameters:

\begin{enumerate}
\item{The cross-scan location of the point source between the two~successive~scans}
\item{The phase, $\psi$, of the point source}
\item{The signal-to-noise ratio of the point source}
\item{The number of repointings between the neighbouring scans}
\item{The roll angle}
\end{enumerate}

By varying the phase, the signal-to-noise ratio and the position of the point source between the two successive scans, the dependence of these parameters on the ability of this method to evaluate the roll angle may be investigated. The results presented here are from 200 noise realisations for every combination of the parameters investigated. Unless otherwise stated the error assumed in the mean spin axis position was 1\arcsec\, and the error in the reference phase was 20\arcmin. This error in the reference phase corresponds to the worst case initial value from the Star Tracker.

\section[]{Results}
\label{section_results}

\subsection{Recovering the roll angle}
\label{section_roll_results}

Two criteria must both be met before this method may be used to evaluate the roll angle,

\begin{enumerate}
\item{The point source is detected at or above the 5$\sigma$ limit in all detectors in both scans}
\item{The point source is found to lie above the detectors in scan 2 and below the detectors in scan 1, as shown in Figure~\ref{roll_fig}.}
\end{enumerate}

These two criteria, together with Tables~\ref{exgal_pred_table} and~\ref{pred_gal_table}, may be used to show that the 545 GHz channel may not be of practical use in extracting the roll angle. The 545 GHz detectors are further from the FRP and are hence more affected by the roll angle than the 857 GHz detectors. For non-zero roll angles the separation of the two neighbouring scans, which meet the constraint of our method that the point source is required to be above all detectors in scan 2 and below all detectors in scan 1, rapidly becomes wider than $6\sigma_b$, where $\sigma_b$ is the $1\sigma$ width of the assumed Gaussian beam. This requires a point source with a flux  $> 120$ Jy in order for the first criterion, the detection of the point source at or above the $5\sigma$ level in all the detectors in both scans, to be met. As seen in Table~\ref{pred_gal_table} there are only 25 galactic point sources $>100$ Jy at 545 GHz, and there are in fact no point sources $>120$ Jy expected at 545 GHz. Given the low numbers of point sources available at 353 GHz, it is therefore expected that the roll angle will be evaluated with data solely from the 857 GHz channel.

\begin{figure}
\begin{center}
\setlength{\unitlength}{1cm}
\begin{picture}(10,9)(0,0)
\put(-1.5,10.5){\includegraphics{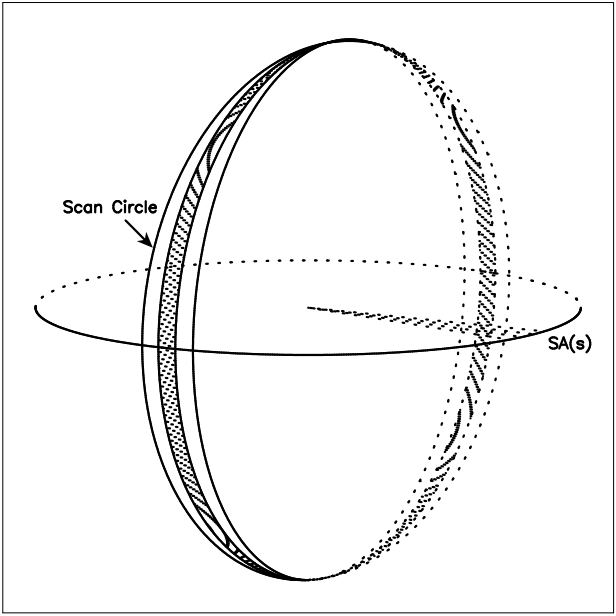}}
\end{picture}
\end{center}
\caption[]{This figure shows a schematic of several neighbouring scans. In the simulated data the point source is constrained to lie on or between two adjacent scans, corresponding to the shaded area in this figure. The roll angle is evaluated using the closest possible two scans in which the requirement that the point source is located above the detectors in scan 2 and below the detectors in scan 1, is met. The point source must also be sufficiently bright to be detected at the $5\sigma$ level in all the detectors on the two scans. It can be seen that the ability to evaluate the roll angle cannot naively be related to the signal-to-noise ratio of the point source, but will also depend on the relative location of the point source to the scan circles and on the roll angle itself.}
\label{area_sphere_fig}
\end{figure}

Using data from the 857 GHz channel, the two successive scans are considered first and if the two criteria are met, then an attempt may be made to reconstruct the roll angle. If the first criterion is met, but the second is not then the next closest scan to the point source is considered and the status of the above criteria are reaccessed. The consideration of further repointings on either side of the shaded area in Figure~\ref{area_sphere_fig}, continues until both criteria are met, and the roll angle may be evaluated, or the first criterion is failed.

\begin{figure}
\begin{center}
\setlength{\unitlength}{1cm}
\begin{picture}(7,7)(0,0)
\put(9,0){\includegraphics{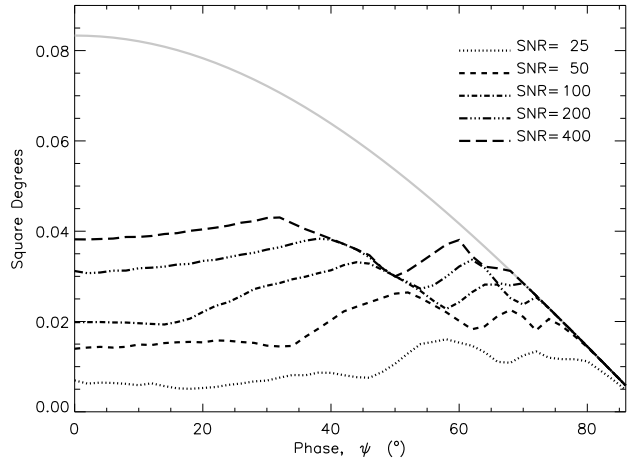}}
\end{picture}
\end{center}
\caption[]{This plot shows the actual area available to the method described here versus the phase of the point source for zero roll angle. The grey curve shows the actual area enclosed by the two scans in each $2\fdg0$ phase bin.}
\label{snr_v_real_area_fig}
\end{figure}
 
\begin{figure}
\begin{center}
\setlength{\unitlength}{1cm}
\begin{picture}(7,7)(0,0)
\put(9,0){\includegraphics{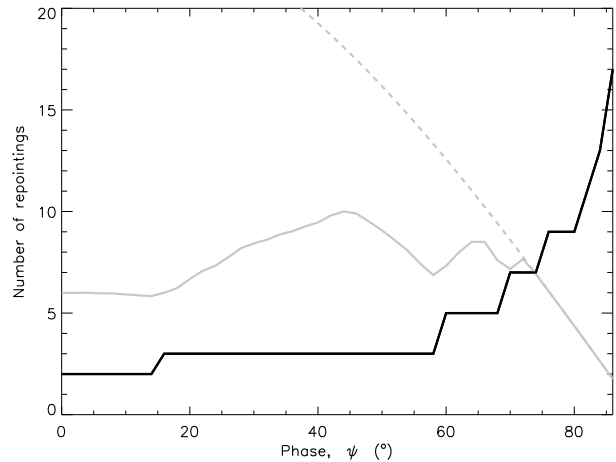}}
\end{picture}
\end{center}
\caption[]{This figure shows the number of repointings between the two scans used versus the phase of the point source for zero roll angle and a 100 signal-to-noise point source. The grey curve, shown for comparison, is the area available to the method for a 100 signal-to-noise point source and zero roll angle as previously shown in Figure~\ref{snr_v_real_area_fig}.}
\label{repoint_area_fig}
\end{figure}

 As the signal-to-noise ratio of the point source increases, the detections will be above the  5$\sigma$ limit in two scans separated by a larger number of repointings. For scans separated by a larger number of repointings the point source is more likely to meet the second criterion. In other words, as the signal-to-noise ratio increases the number of cross-scan and phase locations for the point source, in the shaded region in Figure~\ref{area_sphere_fig}, which allow an attempt to be made in reconstructing the roll angle, increases.  This may be seen in Figure~\ref{snr_v_real_area_fig}, which shows the actual area available to this method versus the phase of the point source for several different signal-to-noise ratios. The shape of the curves in Figure~\ref{snr_v_real_area_fig} are determined by the two criteria which must be met. In reality both criteria may be met for a number of scans, in this case the scans separated by the largest number of pointings will correspond to the greatest coverage of the shaded region in Figure~\ref{area_sphere_fig}. It is this maximum value for the actual area available which is plotted in  Figure~\ref{snr_v_real_area_fig}. The dependence of the curves in Figure~\ref{snr_v_real_area_fig} on the  two criteria may be more easily shown in Figure~\ref{repoint_area_fig}. In this figure the number of repointings, of the nominal 2.5\arcmin, between the two scan circles used is plotted against the phase of the point source. The area available to the method for a 100 signal-to-noise ratio point source is overplotted for comparison. 
 
As the phase increases, the angular distance between the two scan circles decreases, this increases the signal-to-noise of the detections of the point source in each scan. It is this effect which is the cause of the first increase in the area available in Figure~\ref{repoint_area_fig}, which is seen to occur when the number of repointings between the two scans used increases. This is due to the decreasing angular separation between the scans which has allowed the first criterion to be passed by two scans separated by an additional repointing. This additional repointing increases the separation between the detectors on each scan used and thus increases the chance of the second criterion being met. The continued increase in the available area with a constant number of repointings between the scans used is again due to the decreasing angular separation of the scans which increases the area over which the first criterion will be met. The subsequent downturn in the area available is due to the effect of the second criterion, as the separation of the detectors on the two scans is decreasing there are fewer possible locations for the point source which allow it to be positioned above the detectors in scan 2 and below them in scan 1. This process of increasing, then decreasing area available with phase is repeated once the angular separation between the scans decreases enough to allow the first criterion to be passed on two scans separated by a larger number of repointings. This cycle of variation with phase continues until at large phases the total area between two successive scans is available to this method.

\begin{figure}
\begin{center}
\setlength{\unitlength}{1cm}
\begin{picture}(7,7)(0,0)
\put(9,0){\includegraphics{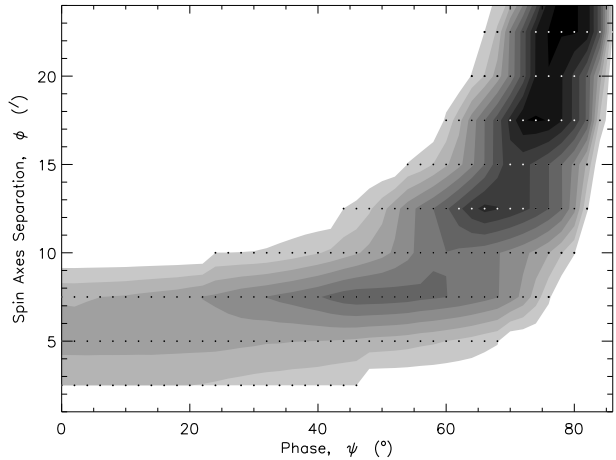}}
\end{picture}
\end{center}
\caption[]{This figure shows a contour plot of the area available to this method given a point source with a signal-to-noise ratio of 100 and zero roll angle. The darker regions correspond to a larger available area. The curves in Figures~\ref{snr_v_real_area_fig} and~\ref{repoint_area_fig} correspond to the maximum value of the available area for a given value of the phase. The black/white dots correspond to the value of phase and spin axis separations used in the simulated data.}
\label{repoint_contour_fig}
\end{figure}

Figure~\ref{repoint_contour_fig} shows a contour plot of the area available, with the darker regions corresponding to a greater available area. The curves in Figures~\ref{snr_v_real_area_fig} and~\ref{repoint_area_fig} correspond to the maximum value of the available area for a given value of the phase. The black and white dots correspond to the value of phase and spin axis separations used in the simulated data. The shape of the contours are defined by the two requirement criteria; the contours towards narrower spin axis separations are governed by the second criterion and the contours towards wider spin axis separations are determined by the first criterion.

\begin{figure}
\begin{center}
\setlength{\unitlength}{1cm}
\begin{picture}(7,7)(0,0)
\put(9,0){\includegraphics{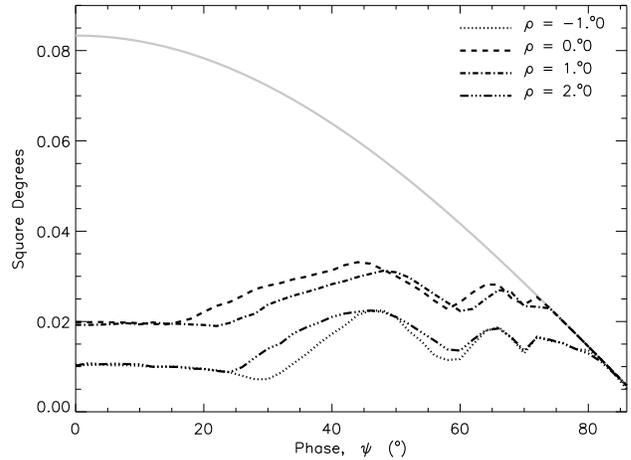}}
\end{picture}
\end{center}
\caption[]{This plot shows the actual area available to the method described here versus the phase of the point source, of a signal-to-noise ratio of 100, for different roll angles. The grey curve shows the actual area enclosed by the two scans in each $2\fdg0$ phase bin}
\label{snr100_v_real_area_fig}
\end{figure}

As the roll angle changes the relative positions of the detectors to the scan, it will also affect whether the second criterion is met. Figure~\ref{snr100_v_real_area_fig} shows the actual area available versus the phase of a point source, of signal-to-noise ratio 100, for several different roll angles. The differences between positive and negative roll angle may be explained by the fact that the positions of the detectors in the focal plane are not symmetric about the FRP. A negative roll angle increases the area covered in the cross-scan direction by the detectors more than a positive roll angle of the same magnitude. The larger the area covered in the cross-scan direction the smaller the chance of the second criterion being met.

\begin{figure}
\begin{center}
\setlength{\unitlength}{1cm}
\begin{picture}(7,7)(0,0)
\put(9,0){\includegraphics{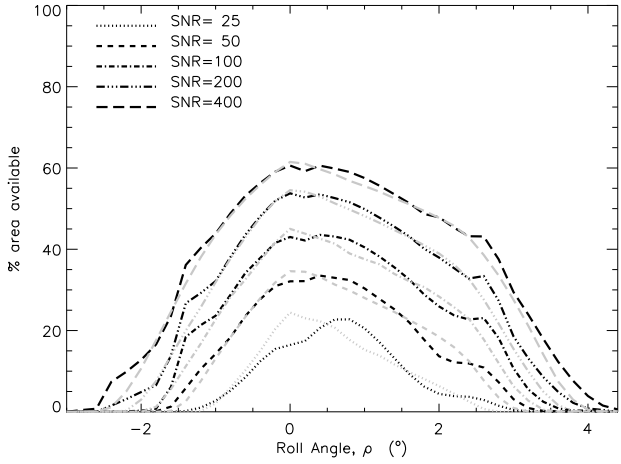}}
\end{picture}
\end{center}
\caption[]{The plot shows how the percentage of the area available varies with the roll angle, for point sources of varying signal-to-noise ratios. The black curves correspond to the number of point sources which are found to meet the two requirement criteria, discussed in the text. Whether the second criterion is met is assessed using a function which determines whether the point source lies above or below the detectors on a scan based on the ratio of their observed amplitudes. The grey curves correspond to the number of point sources included in the simulated data which meet the two requirement criteria.}
\label{roll_v_real_area_fig}
\end{figure}

Figure~\ref{roll_v_real_area_fig} shows how the percentage area of the shaded area in Figure~\ref{area_sphere_fig} which corresponds to the area between two successive scans, varies with the roll angle for point sources of different signal-to-noise ratios. The area available to this method is determined by the requirement criteria; (i) the point source detections in each scan being above the required signal-to-noise ratio and (ii) the location of the point source with respect to the detectors in each of the scans used to extract the roll angle. The grey curves shows how the percentage area varies with roll angle using the input to the simulation to access whether the position of the point source meets the second criterion. The black curves use a routine to evaluate whether the point source lies above or below the scan in question based on the relative amplitudes of the point source as seen by each detector. This routine makes no assumptions on the value of the roll angle for this evaluation. While this will initially be the case, once the roll angle has been evaluated and its time variability modelled, the routine to evaluate whether the second criterion is passed may incorporate the previous values of the roll angle to improve upon its acceptance/rejection of the use of a point source for the roll angle evaluation.

\begin{figure}
\begin{center}
\setlength{\unitlength}{1cm}
\begin{picture}(7,7)(0,0)
\put(9,0){\includegraphics{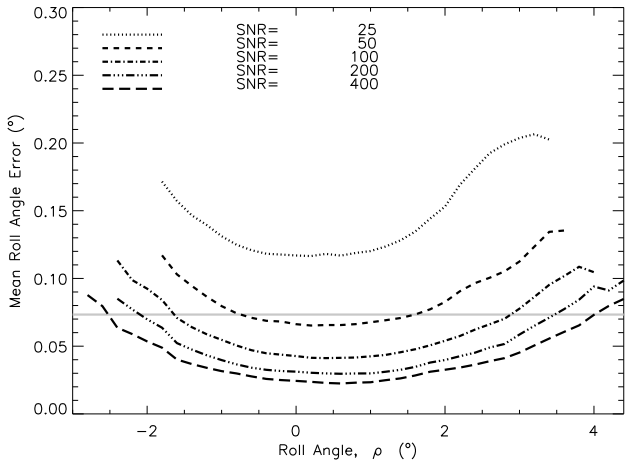}}
\end{picture}
\end{center}
\caption[]{This plot shows the error on the recovered roll angle against the actual roll angle for point sources with different signal-to-noise ratios. The grey line shows the requirement on the roll angle error of 4.4\arcmin.}
\label{snr_v_roll_fig}
\end{figure}

Figure~\ref{snr_v_roll_fig} shows the mean error on the recovered roll angle versus the actual value of the roll angle for point sources of different signal-to-noise ratios, averaged over the 200 realisations of the noise used and the other parameters listed in Section~\ref{section_simData}. The grey line is the requirement on this error as outlined in Section~\ref{section_acc_req}. The error on the recovered roll angle is not dependent on the actual value of the roll angle for small magnitude roll angles, but as the roll angle increases so does the error on the recovered roll angle. Indeed, for larger roll angles it will be difficult to recover the roll angle using this method. This is due not only to the increasing error in the recovery of the roll angle, but also to the decreasing area available to this method with increasing roll angle as seen in Figure~\ref{roll_v_real_area_fig}. This decrease in area available corresponds to a decrease in the number of suitable point sources.

\begin{figure}
\begin{center}
\setlength{\unitlength}{1cm}
\begin{picture}(7,7)(0,0)
\put(9,0){\includegraphics{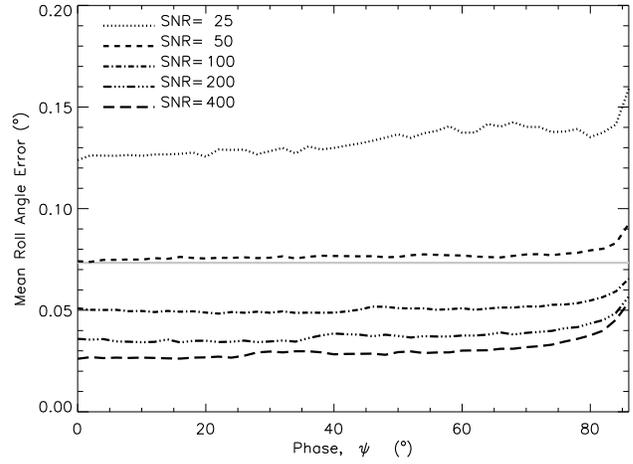}}
\end{picture}
\end{center}
\caption[]{This plot shows the error on the recovered roll angle against the phase location of the point source for varying signal-to-noise ratios. The grey line shows the requirement on the roll angle error of 4.4\arcmin.}
\label{snr_v_phase_fig}
\end{figure}

\begin{figure}
\begin{center}
\setlength{\unitlength}{1cm}
\begin{picture}(7,7)(0,0)
\put(9,0){\includegraphics{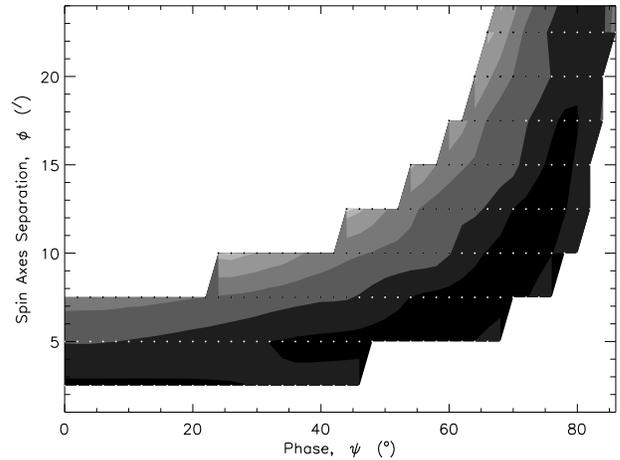}}
\end{picture}
\end{center}
\caption[]{This figure shows contours of constant mean error in the recovered roll angle for zero roll angle and a point source with a signal-to-noise ratio of 100, against the phase of the point source and the angular separation of the spin axes of the scans used. All the contours shown are have values below that of the requirement on the error in the roll angle, as discussed in Section~\ref{section_acc_req}. The darker regions correspond to lower recovered errors on the roll angle. The black/white point correspond to the actual values of phase and spin axis separations used in the simulated data.}
\label{mean_err_contour_fig}
\end{figure}

Figure~\ref{snr_v_phase_fig} shows the mean recovered roll angle error, over the noise realisations and other parameters investigated, for different signal-to-noise ratio point sources versus the location in phase of the point source. The increase in the value of the mean recovered error at large phases is due to the error in the reference phase. This error results in an uncertainty  in the phase of the point source which results in an error in the angular separation, via equation~\ref{s12_eqn}, of the two scans used to recover the roll angle. Given the value used for the error in the reference phase is the expected worst case scenario for this value, the error in the recovered roll angle will have very little dependence on the error in the reference phase.

Figure~\ref{mean_err_contour_fig} shows the dependence of the mean error in the roll angle on the phase, and the angular separation of the spin axes corresponding to the two scan circles used in the evaluation of the roll angle. In this plot a point source with a signal-to-noise ratio of 100 was used to reconstruct the roll angle, which in this case has the value of zero degrees. All the contours in this plot exceed the requirement on the error in the recovered roll angle as discussed in Section~\ref{section_acc_req}, with the darker regions corresponding to lower errors. The black/white points correspond to the actual values of phase and spin axis separations used in the simulated data. This figure shows that optimum phase location for a point source is at intermediate to high phase, with the error increasing again above $\sim$80\dg\, in phase.

\subsection{Recovering the cross-scan position}
\label{section_cs_results}

Using equations~\ref{cross1_eqn} and~\ref{cross2_eqn} from Section~\ref{section_eff_boresight} it is possible to evaluate the cross-scan position of the point source used in evaluating the roll angle as above. This method of evaluating the cross-scan position of the point source is immune to relative calibration errors between the detectors and can incorporate different beams for each of the detectors.

\begin{figure}
\begin{center}
\setlength{\unitlength}{1cm}
\begin{picture}(7,7)(0,0)
\put(9,0){\includegraphics{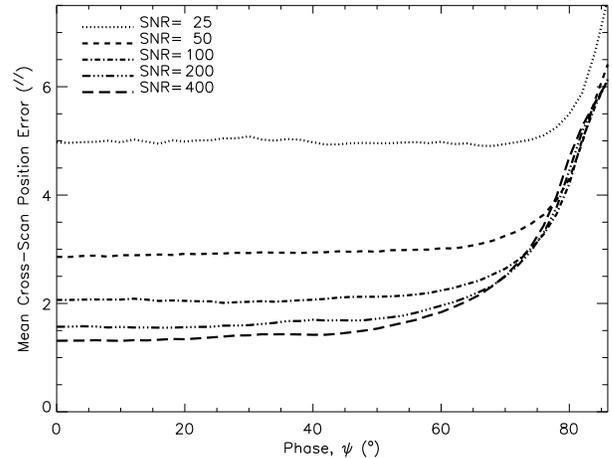}}
\end{picture}
\end{center}
\caption[]{This figure shows the mean error in the recovered cross-scan position versus the location in phase of the point source, for different signal-to-noise ratios.}
\label{cross_scan_err_vs_phase_fig}
\end{figure}

Figure~\ref{cross_scan_err_vs_phase_fig} shows the mean error in the cross-scan position versus the location in phase of the point source, for point sources of different signal-to-noise ratios. The other parameters remaining in the simulated data are averaged over together with the 200 noise realisations. The increase in the error towards larger phases is due to the dependence of the cross-scan position on the angular separation, $S_{12}$, between the two scans at the phase location of the point source. The error in the angular separation of the two scans is dependent on the error in the reference phase; in these simulations the uncertainty in the reference phase is taken to be the worst case scenario value of $20\arcmin$. For more realistic values of the error in the reference phase the error in the cross-scan position at large phases will reduce towards the low phase values.

\begin{figure}
\begin{center}
\setlength{\unitlength}{1cm}
\begin{picture}(7,7)(0,0)
\put(9,0){\includegraphics{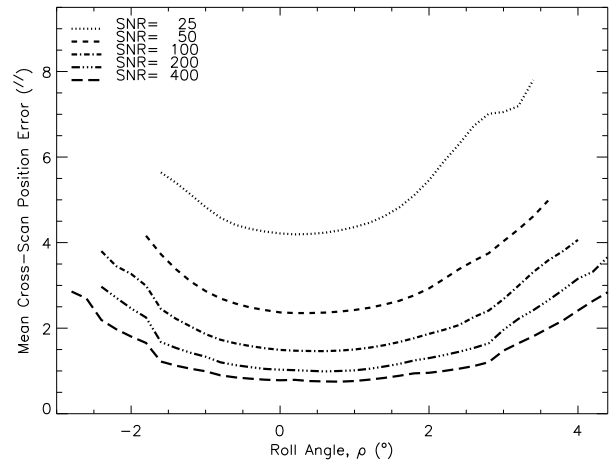}}
\end{picture}
\end{center}
\caption[]{This figure shows the mean error in the recovered cross-scan position versus the roll angle, for different signal-to-noise ratio point sources. The error on the reference phase in this case is 1\arcmin.}
\label{cross_scan_err_vs_roll_fig}
\end{figure}

Figure~\ref{cross_scan_err_vs_roll_fig} shows the mean error in the recovered cross-scan position versus the roll angle, for different signal-to-noise ratio point sources. This figure shows that the error in the recovered cross-scan position follows the same behaviour as the error in the recovered roll angle; no dependence on the actual value of the roll angle for small magnitudes, with an increase in the error towards larger magnitude roll angles. Since the other parameters in the simulated data have been averaged over, the larger errors in the recovered cross-scan position at high phases, as seen in Figure~\ref{cross_scan_err_vs_phase_fig}, will dominate the cross-scan errors as plotted in Figure~\ref{cross_scan_err_vs_roll_fig}. It is for this reason that the error in the reference phase in the case of Figure~\ref{cross_scan_err_vs_roll_fig} has been reduced to 1\arcmin. The overall behaviour of the error in the cross-scan position versus roll angle is the same in both cases.

The area available to this method will be the same as the area available to the roll angle method as described above, since the same criteria on the position of the point source relative to the two scans and its detection thresholds, as described above, must be met.

\subsection{The number of point sources available to the two ring method}
\label{section_pts_available}

Using the predicted numbers of point sources above the given signal-to-noise ratios, as found in Section~\ref{pt_sources_section}, together with the area available to this method found above, it is possible to estimate the number of point source detections with which the roll angle may be evaluated. The area available, for a given roll angle, depends not only on the signal-to-noise ratio of the point source, but also on its location in phase as seen in Figure~\ref{snr_v_real_area_fig}. It is therefore necessary to know the distribution of the point sources in phase before the numbers available to this method may be determined.

\subsubsection{Extragalactic point sources}
\label{section_extgal}

\begin{table}
\caption{SNR and the area available for roll angle, $\rho=0\fdg0$.}
\label{snr_area_table}
\begin{tabular}{|c|c|}
\hline
SNR & \% area available for $\rho=0\fdg0$\\
\hline
50 & 32.11 \\
100 & 43.02 \\
200 & 53.77 \\
400 & 60.60 \\
\hline
\end{tabular}
\end{table}

In the case of extragalactic sources, it may be assumed that the point sources are distributed uniformly throughout the sphere and hence uniformly in phase. The number of sources available to this method therefore, for a given roll angle, may be estimated using Figure~\ref{roll_v_real_area_fig}. The value for percentage area accessible to this method, at zero roll angle, is shown in Table~\ref{snr_area_table}, using these values the number of point sources available to this method may be estimated, and are shown in Table~\ref{exgal_numpts_table}.
\begin{table}
\caption{SNR versus the number of extragalactic point sources available for roll angle, $\rho=0\fdg0$}
\label{exgal_numpts_table}
\begin{tabular}{|c|c|c|}
\hline
SNR & Total number & number available for $\rho=0\fdg0$\\
&  $>$ SNR & $>$ SNR\\
\hline
50 & 59 & 18.7\\
100 &22 &  9.4\\
200 & 9 &  4.8\\
400 & 5 &  3.0\\
\hline
\end{tabular}
\end{table}

The percentage area available decreases with the increasing magnitude of the roll angle, as seen in Figure~\ref{roll_v_real_area_fig}, hence the number of available point sources will also decrease with the increasing roll angle. In the case of zero roll angle the number of extragalactic point sources available is less than one per week, so it is clear that the evaluation of the roll angle will depend predominantly on the galactic and not extragalactic point sources.

\subsubsection{Galactic Point Sources}
\label{section_gal}

A uniform distribution in phase is not a good assumption in the case of galactic point sources, since their distribution in phase will be determined by the position of the galactic plane in terms of phase. This in turn depends upon the scanning strategy used. In order to estimate the numbers of galactic point sources available to the method, assumptions must be made on the distribution of the point sources in the galactic plane and on the scanning strategy employed.  The numbers of galactic point sources available to the method will also depend on the time of year of the observations due to constraints on scanning strategy, determined by the relative locations of the Earth, Sun, and the spin axis position.

In order to investigate these dependencies, the phase coverage of the galactic plane was evaluated on a weekly basis for a given scanning strategy. The number of point sources available to this method may then be found, both on a weekly basis and over the course of the whole mission. It was found that the distribution assumed in galactic latitude had little effect on the number of point sources available, whereas the assumed distribution in galactic longitude had a large effect on the variation of the number of point sources available per week. 

In order to obtain the most reliable estimates of the number of point sources available on a weekly basis; the galactic point sources found in Section~\ref{pt_sources_section}, were binned into 10\dg\, bins in galactic longitude, and within these bins they were assumed to be uniformly distributed between -5\dg\, and +5\dg\, in galactic latitude. 

The first scanning strategy considered, was the baseline strategy of the spin axis remaining in the ecliptic plane throughout the course of the mission, with the nominal 1\dg\, motion in ecliptic longitude per day, maintaining the spin axis in the sunwards direction. Here the boresight angle is defined as the angle between the negative, anti-sunwards direction, spin axis and the line of sight. Table~\ref{basic_ss_roll} shows the number of point source available, for different values of the roll angle, from which the roll angle may be successfully evaluated with an error below the required limit, as discussed in Section~\ref{section_acc_req}. The table shows the minimum, maximum and average number of sources available in the course of one week over the course of the mission together with the total number of point sources available over the whole mission; here the mission is assumed to last for the nominal one year with two sky coverages. Table~\ref{basic_ss_roll} also shows the effect of the decrease in the area available to this method with increasing magnitude of the roll angle, as seen in Figure~\ref{roll_v_real_area_fig}, with the decrease in the number of available point sources for the roll angle values of $\rho = -1\fdg0$ and $\rho = +2\fdg0$.

\begin{table}
\caption{This table displays the number of point sources available, for a given roll angle, $\rho$, assuming that the spin axis remains in the ecliptic plane throughout the mission. The total number of point sources available over the course of the mission, assumes that the sky is covered twice. This table displays the number of point sources available which would provide a measurement of the roll angle with an error less than the requirement of 4.4\arcmin\, as discussed in Section~\ref{section_acc_req}.}
\label{basic_ss_roll}
\begin{tabular}{|l|c|c|c|}
\hline
& $\rho = 0\fdg0$ & $\rho = -1\fdg0$ & $\rho = +2\fdg0$ \\
\hline 
Total number & 355.0 &  169.1 & 288.7 \\
over mission & & &\\
weekly minimum &1.6 & 0.8 & 0.7 \\
weekly maximum &20.8 & 10.8 & 20.1 \\
weekly average &6.9 & 3.3 & 5.6\\
\hline
\end{tabular}
\end{table}

Sinusoidal scanning strategies may also be considered. These scanning strategies may be defined by the path of the spin axis position $(\beta_z,\lambda_z)$, over the course of the mission, given by:
\begin{eqnarray}
\label{scan_strat_eqn}
\lambda_z & = &\lambda,  \nonumber\\
\beta_z & = & A \sin(n\lambda + \phi),
\end{eqnarray} where A is the amplitude of the motion out of the ecliptic plane,  A=0\dg\, is equivalent to the above scanning strategy with the spin axis remaining in the ecliptic plane throughout the whole mission. $\phi$ defines where the spin axis crosses the ecliptic plane and n is the number of periods of the sinusoidal motion over the course of the mission. 

The parameters, A, n and $\phi$ were allowed to vary in order to investigate possible scanning strategies which will maximise the number of point sources available, in particular those scanning strategies with larger weekly minimum values. The value of A is constrained to be $\le$ 10\dg\, as the spin axis must remain within 10\dg\, of the Sun, this constraint is due to the physical size of the solar panels and the need to minimise stray-light.

It was found that the value of $\phi$ had the most effect on the number of point sources available. The values of A and n also have an effect on the number of point sources available, but the values which maximise the number available depend on the value of the roll angle. Given that the value of $\phi$ cannot be predetermined as it too is constrained, by the Earth aspect angle which must remain $\le$ 15\dg\, within the L2 orbit, it is not possible to recommend a particular scanning strategy for the benefit of the roll angle reconstruction.

\begin{table}
\caption{The variation of the weekly minimum value over various sinusoidal scanning strategies, for different roll angles.}
\label{sinusoidal_ss_roll_wk_min}
\begin{tabular}{|r|c|c|c|}
\hline
Roll angle & \multicolumn{3}{c}{Weekly minimum}\\
(\dg) & minimum & maximum & average \\
\hline
-1.0 & 0.4 & 1.4 & 0.8\\
0.0 & 1.1 & 2.1 & 1.6\\
+2.0 & 0.4 & 1.8 & 0.8 \\
\hline
\end{tabular}
\end{table}

\begin{table}
\caption{The variation of the average weekly value over various sinusoidal scanning strategies, for different roll angles.}
\label{sinusoidal_ss_roll_wk_ave}
\begin{tabular}{|r|c|c|c|}
\hline
Roll angle & \multicolumn{3}{c}{Weekly average}\\
(\dg) & minimum & maximum & average \\
\hline
-1.0 & 2.9 & 4.4 & 3.5\\
0.0 & 6.2 & 7.7 & 6.9\\
+2.0 & 4.9 & 6.6 & 5.6 \\
\hline
\end{tabular}
\end{table}

Tables~\ref{sinusoidal_ss_roll_wk_min} and~\ref{sinusoidal_ss_roll_wk_ave} show the variation in the numbers of point sources available over the range of sinusoidal scanning strategies investigated. Table~\ref{sinusoidal_ss_roll_wk_min} shows the variation in the value of the weekly minimum for different values of roll angle, from this table it may be seen that as the magnitude of the roll angle increases so does the likelihood of there being an occasional week in which there will not be a suitable point source available to extract the current value of the roll angle.

\subsection{The impact of the error in the recovery of the effective boresight}
\label{section_error_budget}

The discussion so far has focused on the extraction of the roll angle, assuming that it is allowed to consume the total error budget for the pointing reconstruction. Considerations of the error in the reconstruction of the effective boresight, $\alpha$, must now be included in the error budget. The error in the effective boresight depends on the error in the angle found between the spin axis position and a point source of known position and the error in the cross-scan position found for the point source. Only the error in positions in the cross-scan direction will have an effect on the error in the effective boresight. If it is assumed that the positional errors are the same in both the scan and cross-scan directions then an expression for the error in the effective boresight, $\sigma_{\alpha}$, may be defined as:

\begin{equation}
\label{error_boresight_eqn}
\sigma_{\alpha}^2=\frac{1}{2} \left( \sigma_{pt}^2+ \sigma_{sa}^2 \right) + \sigma^2_{c},
\end{equation}where  $\sigma^2_{c}$ is the error in the cross-scan position of the point source and $\sigma_{pt}$ and $\sigma_{sa}$ are the radial errors in the position of the point source and the spin axis, respectively. 

Table~\ref{error_budget_table} shows the effect of including the errors in the effective boresight on the requirements for the error in the roll angle in order that their combined effect on the pointing error is less than the requirements found in Section~\ref{section_acc_req}. Included in this table are the requirements on the radial error in the point source position if it is assumed that the error in the cross-scan position found for the point source is 3\arcsec, which is the worst case value for a point source with a signal-to-noise ratio of 50 as seen in Figure~\ref{cross_scan_err_vs_phase_fig}.

\begin{table}
\caption{This table shows the requirements on the error in the roll angle, given an error in the effective boresight angle. The error in the effective boresight assumes an error in the cross-scan position of the point source of $\sigma_c$=3\arcsec, an error in the spin axis position of 1\arcsec\, and an error in the point source position as given below. The requirements on the error in the roll angle are given for the two cases of 70\% and 100\% sky coverage.}
\label{error_budget_table}
\begin{tabular}{|c|c|c|c|}
\hline
 $\sigma_{\alpha}$ & \multicolumn{2}{|c|}{$\sigma_{\rho}$} & $\sigma_{pt}$\\
&  $f_{sky}=0.7$ & $f_{sky}=1.0$ & \\
(\arcsec) & (\arcmin) & (\arcmin)&(\arcsec) \\ 
\hline
4 & 4.6 & 4.0 & 3.6\\
5 & 4.4 & 3.7 & 5.6\\
6 & 4.1 & 3.4 & 7.3\\
7 & 3.7 & 3.0 & 8.9\\
\hline
\end{tabular}
\end{table}

\begin{table}
\caption{The variation of the weekly minimum value over various sinusoidal scanning strategies, for different requirements on the roll angle error, for $\rho=0\fdg0$.}
\label{sinusoidal_ss_err_wk_min}
\begin{tabular}{|c|c|c|c|}
\hline
Roll angle & \multicolumn{3}{c}{Weekly minimum}\\
Error $\le$ & minimum & maximum & average \\
\hline
4.4\arcmin & 1.1 & 2.1 & 1.6 \\
4.0\arcmin & 1.0 & 2.2 & 1.5 \\
3.7\arcmin & 1.0 & 1.8 & 1.3 \\
3.4\arcmin & 0.7 & 1.7 & 1.1 \\
3.0\arcmin & 0.6 & 1.5 & 0.9 \\
\hline
\end{tabular}
\end{table}

\begin{table}
\caption{The variation of the weekly average value over various sinusoidal scanning strategies, for different requirements on the roll angle error, for $\rho=0\fdg0$.}
\label{sinusoidal_ss_err_wk_ave}
\begin{tabular}{|c|c|c|c|}
\hline
Roll angle & \multicolumn{3}{c}{Weekly average}\\
Error $\le$ & minimum & maximum & average \\
\hline
4.4\arcmin & 6.2 & 7.7 & 6.9 \\
4.0\arcmin & 5.7 & 7.4 & 6.4 \\
3.7\arcmin & 5.3 & 6.8 & 5.9 \\
3.4\arcmin & 5.0 & 6.2 & 5.5 \\
3.0\arcmin & 4.4 & 5.7 & 4.9 \\
\hline
\end{tabular}
\end{table}

Tables~\ref{sinusoidal_ss_err_wk_min} and~\ref{sinusoidal_ss_err_wk_ave} show the effect of the increased requirements on the accuracy of the roll angle, on the variations in the weekly minimum and average values of the number of available point sources over the various sinusoidal scanning strategies. In Table~\ref{sinusoidal_ss_err_wk_min} the minimum value of the weekly minimum drops below one for the requirement that the error in the roll angle, $\sigma_{\rho}$, is $ \le 3.4$. The average value of the weekly minimum is less than one for the requirement that $\sigma_{\rho}$ is $ \le 3.0$. In this table the value of the roll angle is zero, which we have earlier established is the value of the roll angle with the greatest number of point sources available for its recovery. It is therefore desirable that the limit of the error on the roll angle is as large as possible in order to increase the range in the values of roll angles for which there may be a weekly reconstruction of the roll angle.

\section{Discussion}
\label{section_discussion}

We have shown here, in Section~\ref{subsec_one_ring}, that the roll angle cannot be evaluated using data from a single ring. The cross-scan position of a point source cannot be evaluated before the roll angle has been found as the cross-scan positions of the detectors are unknown. The cross-scan position of a point source can, however, be evaluated using data from a single ring, once the roll angle is known. The cross-scan position of a point source may also be evaluated by an extension to the method used to reconstruct the roll angle from two neighbouring rings, as discussed in Section~\ref{two_rings}. The evaluation of the cross-scan position using two rings is immune to the relative calibration errors between the detectors, but the area available to this method is the same as the area available for the reconstruction of the roll angle; where the area available depends on both the signal-to-noise ratio of the point source and on the actual value of the roll angle. In comparison the area available to the extraction of the cross-scan position from a single ring is only dependent on the value of the roll angle, through its effect on the cross-scan positions of the detectors. In order to extract its cross-scan position, the point source must lie within the range of the cross-scan positions of the detectors. This criterion leads to an opposite dependence on the roll angle, as compared to the two ring method; the area available increases with the magnitude of the roll angle, with the minimum value occuring at zero roll angle. 

The main components of the error in the effective boresight are the error in the recovered cross-scan position and the error in the position of the point source. The error in the effective boresight affects the requirement on the roll angle error in order that the resultant error in the pointing meets the requirements as found in Section~\ref{section_acc_req}. The less stringent the requirements on the roll angle error, the greater the range in the roll angle which may be reliably reconstructed on a weekly basis. It is therefore possible to place requirements on the radial error in the position of the point source used in the evaluation of the effective boresight. Using typical values of the error in the recovered cross-scan position, it is found that in order to meet the pointing requirement, the radial position error should be less than 10\arcsec. Ideally the radial error should be $\sim$3\arcsec\, in order to reduce the requirements on the recovery of the roll angle.

\section{Conclusions}
\label{section_conclusions}

We have investigated methods for the evaluation of parameters required for the pointing reconstruction of the {\it Planck} satellite. The effect of pointing errors on the recovered $C_{\ell}$s has been assessed and requirements on the accuracy of the pointing reconstruction found, such that the effect of random pointing errors is less than the unsubtractable noise. The accuracy at which the roll angle is required is determined so that the uncertainty in the positions of the detectors due to the error in the roll angle is less than the limit on the pointing error.

It is found that it is impractical to determine the roll angle from data from a single ring, as this would require an extremely high signal-to-noise ratio point source to successfully determine the roll angle. An alternative method using two neighbouring rings was developed, which allows the successful extraction of the roll angle to the required accuracy from a single point source, with a signal-to-noise ratio  $\sim 50 $. This method also has the additional advantage of using the ratio of the amplitudes of a point source as seen by the same detector on different rings, hence this method does not require the relative calibration of the detectors. This method, however, does rely upon the assumption that the roll angle is a slowly varying parameter, and it should be remembered that any abrupt changes to the inertia tensor will invalidate this assumption. Once the roll angle is determined the cross scan position of any point source may be assessed. The effective boresight angle may then be evaluated using the cross scan position of a point source of known position.  

Due to the depletion of consumables over the course of the mission and the resultant variation in the values of the pointing parameters, these parameters should be evaluated on a regular basis. Whether a method may be applied  successfully on a regular basis, will depend on the availability of suitable point sources over the course of the mission. In order to investigate this availability, the number of galactic and extragalactic point sources detectable by the {\it Planck} mission was determined. It was found that there is an insufficient number of suitable extragalactic point sources to evaluate these parameters, the roll and effective boresight angles, on a weekly basis. Hence the number of galactic point sources available is crucial. The total number of galactic point sources observable by {\it Planck} over the course of the mission, and more importantly the variation in the number observable on a weekly basis, depends upon the scanning strategy used. However, it was found that it was not possible to recommend a scanning strategy which will  maximise the number of galactic point sources available. This is due to other constraints on the scanning strategy such as maintaining the Earth aspect angle $\le$15\dg\, within the L2 orbit.

The number of point sources available is also found to depend on the value of the roll angle. The method, presented here for the reconstruction of the roll angle from two neighbouring rings, is best suited for the evaluation of small magnitude roll angles. The larger the roll angle, the smaller the number of point sources available and the greater the chance of weeks with no suitable point sources.  For roll angles $\sim +2$\dg\, and $\sim -1$\dg\, it is likely that there will be an occasional week in which there is no suitable point source to evaluate the roll angle. As the magnitude of the roll angle increases above these values the number of suitable point sources diminishes until  $\sim+4$\dg\, and $\sim-3$\dg\, where the evaluation of the roll angle using the method presented here is no longer possible. This places a manufacturing requirement on the alignment of the focal plane with respect to the nominal scan direction.

While the evaluation of the roll angle does not require a known point source, the reconstruction of the effective boresight requires the position of the point source to be known. A known point source catalogue will therefore be required for the evaluation of the effective boresight angles, with positional errors less than 7\arcsec\, and ideally less than 2\arcsec, in order to meet the accuracy requirements on the pointing parameters.

Given the success of this method depends on the numbers of bright point sources available, and for the reconstruction of the effective boresight on their accurate positions, it is important that the IRAS sources which are found here to be potentially suitable for the reconstruction of these parameters are observed prior to launch, to confirm their point source nature, to improve upon their flux estimates and acquire accurate positions. Any point source $\geq 50$ times the noise expected in a binned ring in the 857 GHz channel is potentially useful in evaluating these parameters. Given the expected noise level this corresponds to point sources $\geq 14.2$~Jy at 857 GHz.

The method presented here will allow the weekly monitoring of the roll angle, given the number of point sources determined here, under the expectation of small roll angles and a slowly varying inertia tensor. When combined with a catalogue of accurate point source positions, this method will also allow the weekly determination of the effective boresight angle. It is expected that the determinations of these two angles from the science data will be used to determine their systematic offsets from the equivalent angles determined more accurately by the Star Tracker.

\section*{Acknowledgements}

We would like to thank the following members of the London Planck Anaylsis Centre for helpful conversations: Steve Serjeant, Luis Mendes and Dave Clements. We would also like to thank Dave Clements and Luis Mendes for reading the manuscript. This work was supported by PPARC at both the London and Cambridge Planck Analysis Centres.

\appendix

\label{lastpage}


\begin{thebibliography}{99}


\bibitem[\protect\citeauthoryear{Beichman et. al.}{1988}]{beichman88} Beichman, C.~A., Neugebauer, G., Habing, H.~J., Clegg, P.~E., Chester, T.~J., 1988, Infrared astronomical satellite (IRAS) catalogs and atlases. Volume 1: Explanatory supplement

\bibitem[\protect\citeauthoryear{Dunne et. al.}{2000}]{dunne00} Dunne, L. Eales, S. Edmunds, M., Ivison, R., Alexander, P., Clements, D.~L., 2000, MNRAS, 315, 115

\bibitem[\protect\citeauthoryear{Dunne \& Eales}{2001}]{dunne01} Dunne, L., Eales, S.~A., 2001, MNRAS, 327, 697 

\bibitem[\protect\citeauthoryear{Knox}{1995}]{knox95} Knox, L., 1995, Phys. Rev. D, 52, 4307

\bibitem[\protect\citeauthoryear{Lammare}{2001}]{lamarre01} Lamarre, J.~M., 2001, 1st Science Planck meeting ESTEC, 'Planck HFI instrument -- Design and performances', (available from Livelink {\it http://astro.estec.esa.nl/llink/livelink})

\bibitem[\protect\citeauthoryear{Puget}{2001}]{puget01} private communication J.~L. Puget

\bibitem[\protect\citeauthoryear{Rowan-Robinson et. al.}{1986}]{mrr86} Rowan-Robinson M., Lock, T.~D., Walker, D.~W., Harris, S., 1986, MNRAS, 222, 273

\bibitem[\protect\citeauthoryear{Saunders et. al.}{2000}]{saunders00} Saunders, W., et. al., 2000, MNRAS, 317, 55

\bibitem[\protect\citeauthoryear{van Leeuwen et. al.}{2001}]{leeuwen01} van Leeuwen, F., et. al., 2002, MNRAS, 331, 975

\end{thebibliography}
\end{document}